\documentstyle[12pt]{article}

\begin{document}

\begin{flushright}

IMSc/2019/12/13 

\end{flushright} 

\vspace{2mm}

\vspace{2ex}

\begin{center}

{\large \bf Black hole or Fuzz ball or a Loop quantum star ? }
\\

\vspace{4ex}

{\large \bf Assessing the fate of a massive collapsing star } \\

\vspace{8ex}

{\large  S. Kalyana Rama}

\vspace{3ex}

Institute of Mathematical Sciences, HBNI, C. I. T. Campus, 

\vspace{1ex}

Tharamani, CHENNAI 600 113, India. 

\vspace{2ex}

email: krama@imsc.res.in \\ 

\end{center}

\vspace{6ex}

\centerline{ABSTRACT}

\begin{quote} 

Using the effective equations in our Loop Quantum Cosmology
(LQC) -- inspired models, we resolve the $(n + 2)$ dimensional
black hole singularity. This resolution in the four dimensional
case is same as that given recently by Ashtekar, Olmedo, and
Singh. We then study the fate of a massive collapsing star in
$(n + 2)$ dimensions, focussing on a singularity resolved black
hole, a string theoretic fuzz ball, and a Loop quantum star --
which is a solution to the the effective modified equations for
static spherically symmetric stars we obtained recently using
LQC ideas. We find qualitatively that a massive collapsing star
is likely to become a Loop quantum star : it is likely to have a
macroscopic core region with Planckian densities and pressures
which is surrounded by a low density corona region extending to
the Schwarzschild radius.

\end{quote}

\vspace{2ex}


\vspace{2ex}









\newpage

\vspace{4ex}

\begin{center}

{\bf 1. Introduction} 

\end{center}

\vspace{2ex}

Consider a massive collapsing star which, in general relativity,
would have formed a black hole with a singularity and with a
horizon. It is of interest to understand what the fate of such a
collapsing star would be in a theory of quantum gravity where
black hole singularities are resolved. See, for example,
\cite{k12} -- \cite{k14b} for a discussion of what may be
physically expected from the resolution of black hole
singularities. 

Ten dimensional string theory is one such theory of quantum
gravity. However, to our knowledge, there is no model or any set
of effective equations which may be used to study the
singularity resolutions in detail. Four dimensional Loop quantum
gravity (LQG) is another such theory of quantum gravity. Within
this theory in the framework of Loop Quantum Cosmology (LQC),
several cosmological singularities have been resolved. These
resolutions are also described well by effective equations which
reduce to general relativity equations in the `classical limit'
\cite{b} -- \cite{as}.

Recently, we constructed LQC -- inspired models by generalising
empirically the LQC effective equations to higher dimensions and
studied several cosmological features of these models \cite{k16}
-- \cite{k19}. We generalised the effective equations further so
as to be applicable to static spherically symmetric stars also
\cite{k19b}. We studied them for a constant density star and
found that, for any mass, its pressures remain bounded from
above and that its radius nearly saturates the Buchdahl
bound. For the sake of brevity, in the following, we refer to a
star as a Loop quantum star if its densities and pressures are
bounded from above and if it is a solution to the effective
modified equations for stars given in \cite{k19b}.

Recently, Ashtekar, Olmedo, and Singh (AOS) have given a
resolution of the Schwarzschild black hole singularity using the
effective equations in Loop Quantum Cosmology \cite{aos, aos2}.
The maximum curvature strength then remains Planckian. In light
of such a resolution of the black hole singularities, and in
light of the effective equations which describe well these
resolutions, one may now explore the fate of a massive
collapsing star.

In this paper, we begin this exploration. We first find that,
using the effective equations in our LQC -- inspired models
given in \cite{k16, k19b}, the singularity of an $(n + 2)$
dimensional black hole can be resolved similarly as in
\cite{aos, aos2}.

We then study the fate of a massive collapsing star in $(n + 2)$
dimensions. We consider the region resulting from the
singularity resolution and estimate its size based on physical
expectations. Then, we focus on three possibilities -- a
singularity resolved black hole, a string theoretic fuzz ball,
and a Loop quantum star -- and present a qualitative scenario of
how a collapse may proceed and end.

After considering these possibilities, it appears that a massive
collapsing star will ultimately become a Loop quantum star which
is, in a sense, a hybrid of a black hole and a fuzz ball :
similar to a singularity resolved black hole, it is likely to
have a macroscopic core region where the densities and pressures
are Planckian and, similar to a fuzz ball, the core is likely to
be surrounded by a low density corona region whose size may
extend beyond the Schwarzschild radius and nearly saturate the
Buchdahl bound.

This paper is organised as follows. In section {\bf 2}, we write
down the general relativity equations for the interior of an $(n
+ 2)$ dimensional black hole. In section {\bf 3}, we present the
effective equations in our LQC -- inspired models and the
resolution of an $(n + 2)$ dimensional black hole singularity.
In section {\bf 4}, we estimate the size of the singularity
resolved region based on physical expectations. In section {\bf
5}, we consider the fates of a collapsing star and discuss the
likely scenario. In section {\bf 6}, we present a brief summary
and conclude by mentioning a few topics for further studies.


\vspace{4ex}

\begin{center}

{\bf 2. Interior of an $(n + 2)$ dimensional black hole}

\end{center}

\vspace{2ex} 

In \cite{aos, aos2}, Ashtekar, Olmedo, and Singh (AOS) have
given a resolution of the Schwarzschild black hole singularity
using the effective equations in Loop Quantum Cosmology.
Recently, we constructed LQC -- inspired models by generalising
empirically the LQC effective equations to higher dimensions, to
include arbitrary functions, and to be applicable to static
spherically symmetric stars also \cite{k16, k19b}. In \cite{k17,
k18, k19}, we studied several cosmological features of these
models. It turns out that, using the effective equations in our
LQC -- inspired models and following closely the methods of AOS,
the singularity of an $(n + 2)$ dimensional black hole can also
be resolved similarly as in \cite{aos, aos2}. We now write down,
in this section, the general relativity equations for the
interior of an $(n + 2)$ dimensional black hole. In the next
section, we give effective equations in our LQC -- inspired
models and describe the resolution of an $(n + 2)$ dimensional
black hole singularity.

Let $x^\mu = (t, \; r, \; \theta^a)$ where $a = 1, 2, \cdots ,
n$ be the coordinates of the $(n + 2)$ dimensional spacetime and
let the line element $d s$ be given by
\begin{equation}\label{ds}
d s^2 \; = \; - \; e^{2 \lambda^0} \; d t^2 + e^{2 \lambda} \;
d r^2 + e^{2 \sigma} \; d \Omega_n^2
\end{equation}
where $d \Omega_n$ is the line element on an $n$ dimensional
sphere of unit radius. The general relativity equations are
given, in the standard notation, by
\begin{equation}\label{gr}
R_{\mu \nu} - \frac {g_{\mu \nu}} {2} \; R \; = \; \kappa^2 \;
T_{\mu \nu} \; \; , \; \; \;
\nabla_\mu T^\mu_{\; \; \nu} \; = \; 0 \; \; . 
\end{equation}
Let $T^\mu_{\; \; \nu}$ be diagonal; $( T^0_{\; \; 0}, T^r_{\;
\; r}, T^a_{\; \; a}) = ( - \rho, \Pi, p_a)$ with $p_a = p$ for
all $a \;$; $\; T = \sum_\mu T^\mu_{\; \; \mu} \;$; and, let the
fields depend on $t$ only. Then, equations (\ref{gr}) give
\begin{eqnarray}
\left( 2 \; \lambda_t + (n - 1) \; \sigma_t \right) \; \sigma_t
& = & \frac {2 \; \kappa^2} {n} \; \tilde{\rho} \;
e^{2 \lambda^0} \label{0} \\
& & \nonumber \\
\sigma_{t t} + (n \; \sigma_t + \lambda_t - \lambda^0_t) \;
\sigma_t & = & \kappa^2 \; \left( \tilde{p} - \frac {\tilde{T}}
{n} \right) \; e^{2 \lambda^0} \label{a} \\
& & \nonumber \\
\lambda_{t t} + (n \; \sigma_t + \lambda_t - \lambda^0_t) \;
\lambda_t & = & \kappa^2 \; \left( \tilde{\Pi}
- \frac {\tilde{T}} {n} \right) \; e^{2 \lambda^0} \label{1} \\
& & \nonumber \\
\rho_t + (\rho + \Pi) \; \lambda_t + n \; (\rho + p) \; \sigma_t
& = & 0 \label{c0}
\end{eqnarray}
where the $t-$subscripts denote derivatives with respect to $t
\;$, $\; \tilde{T}^\mu_{\; \; \nu} = T^\mu_{\; \; \nu} +
T^\mu_{\; \; \nu \; *}$ and, as prescribed in \cite{k19b}, we
have treated the curvature terms of the $n$ dimensional sphere
as part of the matter sector and denoted its energy momentum
tensor by $T^\mu_{\; \; \nu \; *} \;$ whose components $(\rho_*,
\Pi_*, p_*)$ are given by
\[
\Pi_* \; = \; - \rho_* \; = \; \frac {n \; (n - 1)}
{2 \; \kappa^2} \; e^{- 2 \sigma} \; \; , \; \; \;
p_* \; = \; \frac {n - 2} {n } \; \Pi_* \; \; .
\]
Hence $T_* = n \; \Pi_*$ and $p_* - \frac {T_*} {n} = \frac {2}
{n} \; \rho_* \;$. For an $(n + 2)$ dimensional black hole,
$\rho = \Pi = p = 0 \;$. Its interior metric is then given by
equation (\ref{ds}) with
\begin{equation}\label{M}
e^{- 2 \lambda^0} \; = \; e^{2 \lambda} \; = \;
\left( \frac {M} {t^{n - 1}} - 1 \right) \; \; , \; \; \;
e^{2 \sigma} \; = \; t^2
\end{equation}
where $0 < t^{n - 1} < M$ and $M \propto \kappa^2 (mass)$ is a
constant. It can be checked easily that equations (\ref{0}) --
(\ref{c0}) are now satisfied.


\vspace{4ex}

\begin{center}

{\bf 3. LQC -- inspired effective equations for the black hole
interior}

\end{center}

\vspace{2ex} 

Following the procedure given empirically in \cite{k16, k19b},
we now present the effective equations for the LQC -- inspired
models. Let $x^\alpha = (x, \; \theta^a)$ be the coordinates of
the $(n + 1)$ dimensional space; let $L^\alpha = (L, \; L^a)$ be
the coordinate lengths of the $\alpha$ -- direction; and let the
line element $d s$ be given by
\begin{equation}\label{dsl}
d s^2 \; = \; - \; d \tau^2 + e^{2 \lambda} \; d x^2
+ \sum_a e^{2 \lambda^a} \; h_a (\theta) \; (L^a d \theta^a)^2
\end{equation}
where $d \tau^2 = e^{2 \lambda^0} d t^2$ and $d \Omega_n^2 =
\sum_a h_a (\theta) (d \theta^a)^2 \;$. Thus
$e^{\lambda^\alpha} L^\alpha$ are the physical lengths of the
$\alpha$ -- direction. Let $c^\alpha = (c, \; c^a)$ denote the
configuration variables and $p_\alpha = (p_c, \; \pi_a)$ their
conjugate momenta. \footnote{ In order to facilitate comparisons
with the results of \cite{aos, aos2}, here and below, we use $b$
and $c$ as configuration variables and as subscripts for the
conjugate momenta. Note that $b$ and $c$ are not to be taken as
spherical indices taking values $1, 2, \cdots, n \;$.} The
variables $p_\alpha$ are given by
\begin{equation}\label{palpha}
p_\alpha \; = \; \frac {V} {e^{\lambda^\alpha} L^\alpha}
\; \; , \; \; \;
V \; = \; \prod_\beta {e^{\lambda^\beta} L^\beta}
\; \; \; \; \Longrightarrow \; \; \; \; 
\prod_\alpha p_\alpha \; = \; V^n \; \; .
\end{equation}
The non vanishing Poisson brackets between $c^\alpha$ and
$p_\alpha$ are given by
\begin{equation}\label{cp}
\{ c^\alpha, \; p_\beta \} \; = \; \delta^\alpha_{\; \; \beta}
\; A \kappa^2 
\end{equation}
where the constant $A$ characterises the $n-$dimensional `area
quantum' : $\lambda_{qm}^n = {\cal O}(1) \; A \kappa^2 \;$.  
Defining $\sigma$ and $p_b$ by
\[
e^{n \sigma} \; = \; \prod_a \left( L^a e^{\lambda^a} \right)
\; \; , \; \; \;
(p_b)^n = \prod_a \pi_a \; \; , 
\]
it follows that 
\begin{equation}\label{cpn}
\{ c^a, \; p_b \} \; = \; \left( \frac {p_b} {\pi_a} \right) \;
\frac {A \kappa^2} {n}  
\end{equation}
and, writing $V $ in terms of $p_\alpha$, that $V^n = (p_b)^n \;
p_c$ and
\begin{equation}\label{lp}
e^\lambda \; = \; \frac {p_b \; p_c^{\frac {1} {n}}} {p_c \; L}
\; \; , \; \; \;
e^{\lambda^a} \; = \; \frac {p_b \; p_c^{\frac {1} {n}}}
{\pi_a \; L^a} \; \; , \; \; \;
e^{n \sigma} \; = \; p_c \; \; .
\end{equation}
After obtaining the equations of motion for $c^\alpha$ and
$p_\alpha$, we will set
\[
(L^a, \; \lambda^a, \; c^a, \; \pi_a) \; = \; (L^\Omega, \;
\sigma, \; b, \; p_b) \; \; \; \forall \; a \; \; .
\]
The line element (\ref{dsl}) then becomes
\begin{equation}\label{dsbar}
d s^2 \; = \; - d \tau^2 \; + \; 
\frac {p_b^2 \; p_c^{\frac {2} {n}}} {p_c^2 \; L^2} \; d x^2
\; + \; p_c^{\frac {2} {n}} \; d \Omega_n^2 \; \; .
\end{equation}


\vspace{4ex}

\begin{center}

{\bf The Hamiltonian and the equations of motion}

\end{center}

\vspace{2ex} 

The equations of motion for $c^\alpha$ and $p_\alpha$ are given
by the `Hamiltonian constraint' $H = 0$ and by the Poisson
brackets of $c^\alpha$ and $p_\alpha$ with $H \;$ : 
\begin{equation}\label{dynamics}
(c^\alpha)_\tau \; = \; \{ c^\alpha, \; H \} \; = \; A \kappa^2
\; \frac {\partial H} {\partial p_\alpha} \; \; , \; \; \;
(p_\alpha)_\tau \; = \; \{ p_\alpha, \; H \} \; = \;
- \; A \kappa^2 \; \frac {\partial H} {\partial c^\alpha}
\end{equation}
where the $\tau-$subscripts denote derivatives with respect to
$\tau \;$. Consider the line element given in equation
(\ref{dsl}) where the lapse function is unity. The corresponding
Hamiltonian $H$ proposed in our LQC -- inspired models
\cite{k16, k19b} is of the form
\begin{equation}\label{H}
H \; = \; H_{grav} (p_\alpha, \; c^\alpha)
+ \tilde{H} (p_\alpha \; ; \; \phi_{mat} , \; \pi_{mat})
\end{equation}
where $\tilde{H} = H_* (p_\alpha) + H_{mat} \;$, the matter
Hamiltonian $H_{mat} = 0 \;$ for black holes, and the
Hamiltonian
\begin{equation}\label{H*}
H_* (p_\alpha) \; = \; V \rho_* \; = \; - \; V \;
\frac {n \; (n - 1)} {2 \; \kappa^2} \; e^{- 2 \sigma} \; = \;
- \; \frac {n \; (n - 1)} {2 \; \kappa^2} \; \left(
p_b \; p_c^{- \frac {1} {n} } \right)
\end{equation}
arises from treating the curvature terms of the $n$ dimensional
sphere as part of the matter sector. The gravity Hamiltonian
$H_{grav}$ is given by
\begin{equation}\label{hgrav}
H_{grav} = - \; \frac {V} {2 A^2 \lambda_{qm}^2 \kappa^2} \;
\sum_{\alpha \beta} G_{\alpha \beta} \; \psi^\alpha \psi^\beta
\end{equation} 
where $G_{\alpha \beta} = 1 - \delta_{\alpha \beta}$ and the
fields $\psi^\alpha$ are given by 
\begin{equation}\label{genpsi} 
\psi^\alpha = \phi^\alpha f^\alpha
\; \; , \; \; \;  \; \; \;
f^\alpha = f(m^\alpha) \; \; , \; \; \;  \; \; \; 
m^\alpha = \bar{\mu}_\alpha c^\alpha
\end{equation}
with the functions $\phi^\alpha, \; f^\alpha$, and
$\bar{\mu}_\alpha$ required to satisfy
\begin{equation}\label{obey}
\phi^\alpha \bar{\mu}_\alpha = \frac {\lambda_{qm} \;
p_\alpha} {V}  \; \; , \; \; \; \; \; \;
f(x) \to x \; \; \; as \; \; \;  x \to 0 
\end{equation}
so that general relativity equations follow in the `classical
limit' $c^\alpha \to 0 \;$. The function $f(x) = sin \; x$ in
LQC; the functions $\phi^\alpha = 1$ for all $\alpha$ in the
$\bar{\mu}-$scheme; and the functions $\mu_\alpha = \mu_0$ for
all $\alpha$ in the $\mu_0-$scheme. In the recent works
\cite{aos, aos2}, AOS take $\mu_\alpha$ to be different
constants : $m^c = \mu_c c$ and $m^a = \mu_b b$ for all $a \;$
where $\mu_c$ and $\mu_b$ are constants.

Following AOS, here we take $f(x) = sin \; x$ and take
$\mu_\alpha$ to be different constants, obtain first the
equations of motion using the Poisson brackets given in equation
(\ref{cp}), and then set $(c^a, \pi_a, \mu_a) = (b, p_b, \mu_b)$
for all $a \;$. However, the same result also follows if one
first sets $(c^a, \pi_a, \mu_a) = (b, p_b, \mu_b)$ for all $a
\;$ in the Hamiltonian $H$ itself but uses the Poisson bracket
given in equation (\ref{cpn}) for $b = c^a$ and $p_b \;$. We
follow the second method. Thus, setting $\psi^a = \psi^b$ for
all $a \;$, we have
\[
\psi^c \; = \; \frac { \lambda_{qm} \; p_c f^c} {V \; \mu_c}
\; \; , \; \; \; 
\psi^b \; = \; \frac { \lambda_{qm} \; p_b f^b} {V \; \mu_b}
\; \; . 
\]
Using $V = p_b p_c^{\frac {1} {n}}$ and $\sum_{\alpha \beta}
G_{\alpha \beta} \; \psi^\alpha \psi^\beta = 2 n \psi^c \psi^b +
n (n - 1) \; (\psi^b)^2 \;$, one gets
\begin{equation}\label{hfinal}
H = - \; \frac {n} {2 A^2 \kappa^2} \; \left( 2 \;
\frac {f^c f^b} {\mu_c \mu_b} \; p_c^{\frac {n - 1} {n}}
+ (n - 1) \; \left( \left( \frac {f^b} {\mu_b} \right)^2
+ A^2 \right) \; p_b \; p_c^{- \frac {1} {n}} \right) \; \; .
\end{equation}
Let $d \tau = N d T$ where $N$ is a lapse function.  Choosing $N
= \frac {A \; \mu_b \; p_c^{\frac {1} {n}}} {f^b} \;$, the
corresponding Hamiltonian $h = N H$ is given by
\begin{equation}\label{h}
h \; = \; - \; \frac {n} {2 A \kappa^2} \; \left( 2 \; \frac
{f^c} {\mu_c} \; p_c + (n - 1) \; \left( \frac {f^b} {\mu_b}
+ \frac {A^2 \mu_b} {f^b} \right) \; p_b \right) \; \; .
\end{equation}
Defining $g_\alpha = \frac {d \; f^\alpha} {d \; m^\alpha} =
\frac {d \; f (m^\alpha)} {d \; m^\alpha} \;$ and with the
$T-$subscripts denoting derivatives with respect to $T \;$, the
equations of motion are now given by
\begin{eqnarray}
c_T \; = \; - \; n \; \frac {f^c} {\mu_c} & , & 
b_T \; = \; - \; \frac {n - 1} {2} \; \left( \frac {f^b} {\mu_b}
+ \frac {A^2 \mu_b} {f^b} \right) \label{cb} \\
& & \nonumber \\
(p_c)_T \; = \; n \; g_c \; p_c & , & 
(p_b)_T \; = \; \frac {n - 1} {2} \; g_b \; \left( 1 - \frac
{A^2 \; (\mu_b)^2} {(f^b)^2} \right) \; p_b \label{pcpb} 
\end{eqnarray}
which, together with $h = 0 \;$, implies that 
\begin{equation}\label{h0}
2 \; \frac {f^c} {\mu_c} \; p_c \; = \; - \; (n - 1) \; \left(
\frac {f^b} {\mu_b} + \frac {A^2 \mu_b} {f^b} \right) \; p_b
\; = \; const \; \; .
\end{equation}
In the classical limit, $c^\alpha \to 0 \;$, hence $f^c \to
\mu_c c$ and $f^b \to \mu_b b \;$. Then $N = \frac {A \;
p_c^{\frac {1} {n}}} {b}$ and equations (\ref{h}) -- (\ref{h0})
give
\begin{equation}\label{clh}
h \; = \; - \; \frac {n} {2 A \kappa^2} \; \left( 2 \; c \; p_c
+ (n - 1) \; \left( b + \frac {A^2} {b} \right) \; p_b \right)
\end{equation}
\begin{eqnarray}
c_T \; = \; - \; n \; c & , & b_T \; = \; - \; \frac {n - 1} {2}
\; \left( b + \frac {A^2} {b} \right) \label{clcb} \\
& & \nonumber \\
(p_c)_T \; = \; n \; p_c & , & (p_b)_T \; = \; \frac {n - 1} {2}
\; \left( 1 - \frac {A^2} {b^2} \right) \; p_b \label{clpcpb}
\end{eqnarray}
\begin{equation}\label{clh0}
2 \; c \; p_c \; = \; - \; (n - 1) \; \left( b
+ \frac {A^2} {b} \right) \; p_b \; = \; const \; \; .
\end{equation}


\vspace{4ex}

\begin{center}

{\bf Solutions}

\end{center}

\vspace{2ex} 

Let the horizon be located at $T = 0 \;$ where $b = b_0 = 0$ and
$p_c = p_{c 0} = M^{\frac {n} {n - 1}} \;$, and let $T$ be $ < 0
\;$ in the interior of the black hole. Equations (\ref{clh}) --
(\ref{clh0}) and, for $f(x) = sin \; x \;$, equations (\ref{h})
-- (\ref{h0}) can be solved explicitly. Solution to equations
(\ref{clh}) -- (\ref{clh0}) are given by

\begin{eqnarray}
c \; = \; c_0 \; e^{- n T} & , & 
b \; = \; A \; \sqrt{e^{- (n - 1) T} - 1} \label{clcbT} \\
& & \nonumber \\
p_c \; = \; M^{\frac {n} {n - 1}} \; e^{n T} & , & 
p_b \; = \; - \; p_{b 0} \; e^{(n - 1) T} \;
\sqrt{e^{- (n - 1) T} - 1} \label{clpcpbT}
\end{eqnarray}
where $2 c_0 p_{c 0} = (n - 1) A \; p_{b 0} \;$. In the line
element in equation (\ref{dsbar}), one has
\[
N^2 \; = \; \frac {M^{\frac {2 n} {n - 1}} \; e^{2 T}}
{e^{- (n - 1) T} - 1} \; \; , \; \; \;
\frac {p_b^2 \; p_c^{\frac {2} {n}}} {p_c^2 \; L^2} \; = \;
\frac {(p_{b 0})^2 \; \left( e^{- (n - 1) T} - 1 \right) }
{(L M)^2 }
\]
Setting $p_{b 0} = L M$ and $t = p_c^{\frac {1} {n}} = M^{\frac
{1} {n - 1}} \; e^T \;$, one obtains the classical interior
metric of the black hole, see equations (\ref{ds}) and
(\ref{M}). Solutions to equations (\ref{h}) -- (\ref{h0}) may be
written, after some straightforward algebra, as
\begin{eqnarray}
& & tan \; \chi(c) \; = \; (tan \; \chi_0) \; e^{- n T}
\; \; , \; \; \; 
cos (\mu_b b) \; = \; B \; tanh \; \zeta(T) \label{cbT} \\
& & \nonumber \\
& & p_c \; = \; M^{\frac {n} {n - 1}} \; \left( cos^2 \chi_0 \;
e^{n T} \; + \; sin^2 \chi_0 \; e^{- n T} \right) \label{pcT} \\
& & \nonumber \\
& & p_b \; = \; - \; \frac {A L M \; \mu_b \; sin (\mu_b b)}
{sin^2 (\mu_b b) + (\mu_b A)^2} \label{pbT}
\end{eqnarray}
where $\chi(c), \; \chi_0, \; B$, and $\zeta(T)$ are defined by
\begin{eqnarray}
\chi(c) \; = \; \frac {\mu_c \; c} {2} & , &
\zeta(T) \; = \; \frac {(n - 1) \; B} {2} \; T + \zeta_0
\label{chi} \\
& & \nonumber \\
\chi_0 \; = \; \frac {\mu_c \; c_0} {2} &, & Coth \; \zeta_0
\; = \; B \; = \; \sqrt{1 + \mu_b^2 \; A^2} \; \; , \label{chi0}
\end{eqnarray}
with $c_0$ now given by $2 \; sin (\mu_c c_0) \; p_{c 0} = (n -
1) \; \mu_c \; A L M \;$. Note that when $e^{n T} = (tan \;
\chi_0)$, the field $p_c$ and the areal radius $t = p_c^{\frac
{1} {n}}$ reach their minima given by
\[
p_{c (min)} \; = \; t_{min}^n \; = \; sin (\mu_c c_0) \; p_{c 0}
\; = \; \frac {n - 1} {2} \; \mu_c \; A L M \; .
\]
Note also that when $\mu_b \; b = \frac {\pi} {2} \;$ which
occurs at $T = - \; \frac {2 \; \zeta_0} {(n - 1) \; B} \;$, the
field $\vert p_b \vert $ reaches its maximum given by
\[
\vert p_b \vert_{max} \; = \; \frac {\mu_b \; A L M}
{1 + (\mu_b A)^2} \; \simeq \; \mu_b \; A L M \; \; .
\]
These minima and maxima occur generically at different times.
Near these times, ignoring factors of ${\cal O} (1)$ and using
equations (\ref{dsl}) and (\ref{dsbar}), we write
\begin{eqnarray}
p_c \; \simeq \; t_{min}^n \; \simeq \; \mu_c L \; M
& \longrightarrow & \mu_c L \; \simeq \; \frac {t_{min}^n} {M}
\label{mucl} \\
& & \nonumber \\
e^{2 \lambda} \; \simeq \; \left( \frac {p_b^2} {L^2} \;
p_c^{\frac {2 (1 - n)} {n}} \right)_{max} \; \simeq \; \frac {M}
{t_{min}^{n - 1}} & \longrightarrow & \mu_b^2 \; \simeq \; \frac
{t_{min}^{n - 1}} {M} \; \; . \label{mub2}
\end{eqnarray}

With additional physical input on $t_{min}$, see section {\bf
4}, equations (\ref{mucl}) and (\ref{mub2}) may be used to
express $(\mu_c L)$ and $\mu_b$ in terms of $M \;$. For this
purpose, noting that $M \propto \kappa^2 (mass)$ and ignoring
factors of ${\cal O}(1)$, let $M \simeq {\cal N} \; l_{pl}^{n -
1} \;$ and $t_{min} \simeq {\cal N}^s \; l_{pl}$ where ${\cal
N}$ is the mass of the black hole in Planck units and the
exponent $s \ge 0$ is a constant. Then equations (\ref{mucl})
and (\ref{mub2}) give
\begin{equation}\label{mucbs}
\mu_c \; L \; \simeq \; {\cal N}^{s n - 1} \; l_{pl}
\; \; , \; \; \;
\mu_b^2 \; \simeq \; {\cal N}^{s (n - 1) - 1} \; \; . 
\end{equation}
Also, a measure of the maximum curvature strength may be given
by
\begin{equation}\label{ries}
\left( R_{a b c d} \right)_{max} \; = \; \left( e^\mu_a \;
e^\nu_b \; e^\lambda_c \; e^\sigma_d \; R_{\mu \nu \lambda
\sigma} \right)_{max} \; \simeq \; \frac {M} {t_{min}^{n + 1}}
\; \simeq \; \frac {{\cal N}^{1 - s (n + 1)}} {l_{pl}^2}
\; \; .
\end{equation}


\vspace{4ex}

\begin{center}

{\bf 4. Size of the resolved region : physical expectations}

\end{center}

\vspace{2ex} 


In a theory of quantum gravity, a black hole singularity is
expected to be resolved. Let the physical size of the region
which results from the singularity resolution be denoted, upto
factors of ${\cal O}(1) \;$, by
\begin{equation}\label{rres}
R_{res} \; \simeq \; {\cal N}^s \; l_{pl} \; \; , \; \; \;
{\cal N} \; \simeq \; \frac {M} {l_{pl}^{n - 1}}
\end{equation}
where ${\cal N}$ is the mass of the black hole in Planck units
and the exponent $s$ is a constant. The size $R_{res} \;$,
equivalently the exponent $s \;$, can be estimated based on what
is physically expected from the resolution of a black hole
singularity. See, for example, \cite{k12} -- \cite{k14b}.

One expects $R_{res} > l_{pl} \;$ which implies that $s \ge 0
\;$. If one expects to be able to describe unitarily the
evolution of the collapsed constituents of the original massive
star then $R_{res}$ needs to be $ \gg l_{pl}$ parametrically
which implies that $s > 0 \;$. At its largest, $R_{res}$ may be
comparable, as in Mathur's fuzzball proposal, to the horizon
size $R_h \simeq {\cal N}^{\frac {1} {n - 1}} \; l_{pl}$ which
implies that $s \stackrel {<} {_\sim} \frac {1} {n - 1} \;$.
Note that, for any $s > 0 \;$, the size $R_{res}$ is macroscopic
since it is $\gg l_{pl}$ parametrically, and that it depends on
the mass of the original star.

When a neutron star forms, the original star's atoms get crushed
to nuclear densities at the center and, through complex
processes, transform into a soup of quarks, gluons, hadrons, et
cetera. Similarly, in a theory of quantum gravity which resolves
the black hole singularity, one may expect the original star's
constituents to get crushed to Planckian densities at the center
and, through complex processes, transform into a soup of
Planckian mass units and, perhaps, other types of less massive
units also. Let ${\cal M} = {\cal N}^\nu$ be the number of such
fundamental units. If these units are of Planckian mass then
$\nu = 1 \;$. If these units are of sub Planckian mass then $\nu
\stackrel {>} {_\sim} 1 \;$. If these units are expected to
describe the Bekenstein entropy $S_{bek} \simeq {\cal N}^{\frac
{n} {n - 1}}$ of the black hole that would have formed in
general relativity then ${\cal M} \simeq S_{bek} \;$, hence $\nu
= \frac {n} {n - 1} \;$. If one is considering the singularity
resolution in an `old black hole', then one expects $\nu
\stackrel {>} {_\sim} \frac {n} {n - 1}$ in order to account for
past accretions and evaporations, for the in-fallen quanta of
Hawking radiation, et cetera \cite{k12} -- \cite{k14b}.

Given ${\cal M}$ fundamental units of the quantum gravity
theory, let the physical size occupied by them be $\simeq {\cal
M}^\delta \; l_{pl} \;$. Then $R_{res} \simeq {\cal N}^{\nu
\delta} \; l_{pl}$, hence $s = \nu \; \delta \;$. If,
conservatively, one takes these units to be maximally densely
packed then $\delta = \frac {1} {n + 1} \;$. If one takes these
units to `random walk' then, perhaps, $\delta \simeq \frac {1}
{2} \;$. Thus it follows that, conservatively,
\begin{equation}\label{s}
\nu \; \ge \; 1 \; \; , \; \; \;
\delta \; \ge \; \frac {1} {n + 1} \; \; , \; \; \;
s = \nu \; \delta \; \ge \; \frac {1} {n + 1} \; \; .
\end{equation}
For $s = \frac {1} {n + 1} \;$, equations (\ref{mucbs}) and
(\ref{ries}) give
\begin{equation}\label{aoss}
\mu_c \; L \; \simeq \; \frac {l_{pl}}
{{\cal N}^{\frac {1} {n + 1}}}
\; \; , \; \; \;
\mu_b^2 \; \simeq \; \frac {1} {{\cal N}^{\frac {2} {n + 1}}}
\; \; , \; \; \; \left( R_{a b c d} \right)_{max}
\; \simeq \; \frac {1} {l_{pl}^2}
\end{equation}
which, for $n = 2 \;$, give the results of AOS in \cite{aos,
aos2}. Also, if $\mu_c$ and $\mu_b$ are fixed as in equations
(\ref{mucbs}) then the maximum curvature strength is sub
Planckian for $s > \frac {1} {n + 1} \;$ : $\; \left( R_{a b c
d} \right)_{max} \simeq \; \frac {{\cal N}^{(- ve)}} {l_{pl}^2}
\ll \frac {1} {l_{pl}^2} \;$.

\newpage

\vspace{4ex}

\begin{center}

{\bf 5. Fate of a massive collapsing star} 

\end{center}

\vspace{2ex} 

Consider a massive collapsing star which, in general relativity,
would have formed a black hole with a singularity and with a
horizon. Consider its possible fates now in light of {\bf (i)}
the LQC resolution of black hole singularities in four
dimensions given in \cite{aos, aos2} and in $(n + 2)$ dimensions
given here; {\bf (ii)} the physical expectations described in
section {\bf 4}; and {\bf (iii)} the effective modified
equations for static spherically symmetric stars obtained in
\cite{k19b} using LQC ideas.

We discuss three possible fates for a massive collapsing star :
{\bf (1)} a black hole with its singularity resolved but with a
horizon present; {\bf (2)} string theoretic fuzz ball proposed
by Mathur which has no horizon and (assumed implicitly to have)
no singularities; and {\bf (3)} a Loop quantum star whose
densities and pressures are bounded from above and which is a
solution to the effective modified equations for stars given in
\cite{k19b}.

\vspace{2ex} 

\centerline{\bf Black hole}

\vspace{2ex}

By construction, the effective LQC equations reduce to general
relativity ones when the spacetime curvature is small. Hence,
the collapsing star will form a blackhole with a horizon but now
with its singularity resolved as given in \cite{aos, aos2} for
four dimensions, and in here for $(n + 2)$ dimensions with $s =
\frac {1} {n + 1} \;$. The singularity is now resolved into a
region of Planckian curvature whose physical size is $\simeq
{\cal N}^{\frac {1} {n + 1}} \; l_{pl} \;$, is macroscopic, and
depends on the mass of the original star.

Thus, as a star collapses, a horizon will form. Its atoms will
be crushed at the centre into Planckian mass objects which
become densely packed. In the horizon region, the black hole
will evolve as in general relativity : will emit Hawking
radiation, evaporate, and shrink. Over a time of order ${\cal
N}^{n + 1} \; t_{pl} \;$, the horizon size will become $\simeq
{\cal N}^{\frac {1} {n + 1}} \; l_{pl} \;$. The evaporation
process then is likely to be modified significantly and
information is likely to emerge.

One then has a macroscopic remnant with Planckian curvature and
with a size which depends on the mass of the original star. It
will evolve unitarily and, depending on the details of the
quantum gravity theory, may remain stable or shrink further.

However, in the LQC and the LQC -- inspired approaches, the
microscopic origin of Bekenstein entropy is not clear. Also, if
matter is present in the black hole interior -- as when the star
has just collapsed to within its Schwarzschild radius or if
there is matter accretion after a black hole is formed -- then,
as mentioned in \cite{aos, aos2}, the horizon may be replaced by
a curvature singularity which gets resolved by effective LQC
equations in the $\bar{\mu}-$scheme.

\vspace{2ex} 

\centerline{\bf Fuzz ball}

\vspace{2ex} 

The collapsing star may form a string theoretic fuzz ball
proposed by Mathur which, by definition, has no horizon. String
theory effects are implicitly assumed to resolve the
singularities.

To our knowledge, there is no model or any set of effective
equations which may be used to study the singularity resolutions
or the pressure and density distributions of the `fuzz' inside a
fuzz ball star. In the absence of such a model or a set of
equations, one may only speculate. Thus, speculatively, a
horizon may form first as a star collapses. The star's atoms get
crushed at the centre into stringy objects which may then spread
out as fuzz upto and beyond the Schwarzschild radius. With a
horizon thus absent, the subsequent evolution will be unitary
just as for a piece of burning coal.

\vspace{2ex} 

\centerline{\bf Loop quantum star}

\vspace{2ex} 

Recently, using LQC ideas, we have obtained effective modified
equations for static spherically symmetric stars
\cite{k19b}. These modifications are in the $\bar{\mu}-$ scheme
and involve one arbitrary function which, when chosen
appropriately, bounds the densities and pressures from above. A
Loop quantum star is a solution to these modified equations
with its densities and pressures bounded from above. 

Inferring from the study of the modified equations for a
constant density star given in \cite{k19b}, a Loop quantum star
is likely to have a core region of size $\simeq {\cal N}^{\frac
{1} {n + 1}} \; l_{pl} \;$, where the densities and pressures
are Planckian, surrounded by a low density corona region which
may extend beyond the Schwarzschild radius and nearly saturate
the Buchdahl bound.

The formation of a Loop quantum star is likely to be a hybrid of
black hole and fuzz ball formation : As a star collapses, a
horizon will form. Its atoms will be crushed at the centre
mostly into Planckian mass objects which become densely packed,
and partly into light coronal objects which may spread out to a
size of Schwarzschild radius. Such low density objects are
perhaps also necessary to describe Bekenstein entropy.

Note that, unlike the LQC case, the exponent $s$ may now be $ >
\frac {1} {n + 1}$ for a Loop quantum star, see section {\bf 4}.
Also, unlike the fuzz ball case, one now has effective modified
equations for such a star which may be solved to obtain static
configurations. Thus, for example, one may explore whether a
core and coronal configuration as described here is possible for
any class of equations of state and for any class of functions.


\vspace{2ex} 

\centerline{\bf Discussion}

\vspace{2ex} 

In all the above cases, when a massive star just collapses to
around its Schwarzschild radius, general relativity should
continue to be applicable and the star's atoms and other
constituents should continue to be described by standard model
physics. Given this, it is inconceiveable that quantum gravity
effects will come into play immediately and halt the collapse.
Hence, the star should develop a horizon as in general
relativity and continue to collapse until its energies,
densities, and pressures reach Planckian values. Quantum gravity
effects may then become operative and dictate the further
evolution of the star into a black hole or a fuzz ball or a Loop
quantum star.

It is important to estimate the time elapsed from when the star
collapses to around its Schwarzschild radius to when the
Planckian objects become visible into which the star's original
constituents have transformed by quantum gravity effects. For a
black hole, this time is of order of its evaporation time ${\cal
N}^{n + 1} \; t_{pl} \;$. For a fuzz ball and a Loop quantum
star, it is possible that this time is of the order of transit
time across a Schwarzschild radius multiplied by a red shift
factor at the Buchdahl limit. But this may be a vast
underestimate and we do not know a more realistic estimate of
this time.

In summary, note that a fuzz ball is qualitatively similar to a
Loop quantum star. Note further that the Planckian curvature
region in a black hole, into which the star's constituents must
have transformed, seems not to have enough fundamental units to
explain Bekenstein entropy. Also, if matter is present in the
black hole interior then the horizon may become singular and get
resolved by LQC effects. This will then render a black hole to
be qualitatively similar to a Loop quantum star. Thus it appears
that a massive collapsing star will ultimately become a Loop
quantum star.

\newpage

\vspace{4ex}

\begin{center}

{\bf 6. Conclusion}

\end{center}

\vspace{2ex} 

We now give a brief summary of the present paper and conclude by
mentioning a few topics for further studies.

We studied the interior of $(n + 2)$ dimensinal black holes
using our LQC -- inspired models. We presented the effective
equations for the interior and obtained the solutions which, for
$n = 2 \;$, reduce to those given by AOS. We then considered a
massive collapsing star which, in general relativity, would have
formed a black hole. We analysed its possible fates in a quantum
theory of gravity which resolves the black hole singularities.
We discussed the singularity resolved black holes in LQC, string
theoretic fuzz balls, and Loop quantum stars as possible
fates. It appears from our discussions that a massive collapsing
star will ultimately become a Loop quantum star.

We now conclude by mentioning a few topics for further studies.

It is important to obtain solutions to the effective modified
equations for a static spherically symmetric star and explore
whether a core and coronal configuration described in this paper
is possible for any class of equations of state and for any
class of functions.

It is desireable to obtain effective equations where the fields
depend on time and a spatial coordinate, and which resolve the
singularities. Such effective equations may then be used to
study the dynamics of the collapsing stars. One may derive such
effective equations rigorously, or obtain them empirically,
based on ideas in Loop Quantum Gravity or in string theory.

The effective equations in LQC and in the LQC -- inspired models
are designed to reduce to general relativity equations when the
spacetime curvature is weak. See, however, the recent comments
of \cite{brahma, b19, bmm} on \cite{aos, aos2}. It is desireable
to obtain effective equations which may be trusted even if their
predictions differ from general relativity's in weak curvature
regions. One may then use such equations to obtain insights
into, for example, Mathur's fuzz ball proposal according to
which no horizon is present even for massive stars.




\vspace{4ex}

{\bf Note Added:} S. Brahma has pointed out to us an alternate
fate of black hole singularities in LQG where the holonomy
corrections responsible for the singularity resolution lead,
upon respecting covariance, to changing of the spacetime
signature from Lorentzian to Euclidean in the high curvature
regime. See \cite{br, bb, bby}.


\end{document}